\documentclass{ifacconf}
\usepackage{natbib}
\usepackage{amsmath}
\usepackage{amsfonts}
\usepackage{mathrsfs}
\usepackage{dsfont}
\usepackage{graphicx}

\newtheorem{defi}[thm]{Definition}

\begin{document}
\begin{frontmatter}
\title{Investigation of Alternative Measures for Mutual Information}


\author[First]{Bulut Kuskonmaz} 
\author[First]{Jaron S. Gundersen} 
\author[First]{Rafal Wisniewski}

\address[First]{Department of Electronical Systems, Aalborg University, Aalborg, Denmark (e-mails: $\{$bku,jaron,raf$\}$@es.aau.dk).}

\begin{abstract}                
Mutual information $I(X;Y)$ is a useful definition in information theory to estimate how much information the random variable $Y$ holds about the random variable $X$. One way to define the mutual information is by comparing the joint distribution of $X$ and $Y$ with the product of the marginals through the KL-divergence. If the two distributions are close to each other there will be almost no leakage of $X$ from $Y$ since the two variables are close to being independent. In the discrete setting the mutual information has the nice interpretation of how many bits $Y$ reveals about $X$ and if $I(X;Y)=H(X)$ (the Shannon entropy of $X$) then $X$ is completely revealed. However, in the continuous case we do not have the same reasoning. For instance the mutual information can be infinite in the continuous case. This fact enables us to try different metrics or divergences to define the mutual information. In this paper, we are evaluating different metrics or divergences such as Kullback-Leibler (KL) divergence, Wasserstein distance, Jensen-Shannon divergence and total variation distance to form alternatives to the mutual information in the continuous case. We deploy different methods to estimate or bound these metrics and divergences and evaluate their performances. 

\end{abstract}


\end{frontmatter}

\section{Introduction}
Mutual information has been used as a measure of privacy leakage in several contexts. It dates back to the introduction of information theory in \cite{S48} where it was introduced through the Shannon entropy. Afterwards, it has been used to measure leakage in several contexts such as multiparty computation, differential privacy, and machine learning \cite{Leak20, FA20, Poor13, Urrutia2018InformationTF, DP_MI16, Li_Jaron21}. In this paper we focus mainly on the use of mutual information in privacy-preserving distributed computations, such as for instance multiparty computation. Even though you want to compute on real numbers, a multiparty computation protocol is usually converted into finite field operations, and hence the mutual information is between two discrete random variables. However, \cite{Katrine_Shamir} suggest secret sharing scheme over the real numbers. With this secret sharing scheme the multiparty computation can be carried out directly in the real numbers which gives the advantage that some real number computations (for example division) are easier to carry out. The drawback of this is that a share might leak a small amount of information, but as described in the paper the amount can be very limited. In any case, this gives rise to study and investigate different ways to measure the leakage in the continuous case as well. We give more details about how mutual information is used to measure leakage in multiparty computation in Section \ref{MI_leak}.

The interpretation of the mutual information $I(X;Y)$ in the discrete case is how many bits $Y$ on average reveals about $X$ and hence if we want to keep $X$ private we want this mutual information to be small if someone can/will get access to $Y$. However, this bit interpretation goes out of the window when the random variables are continuous random variables (since we need an infinite number of bits to describe a real number). Therefore, in the continuous case there might be other alternatives which are just as good as mutual information to measure information leakage. For instance, if we define the mutual information through the KL divergence we measure the difference between the joint distribution of $X$ and $Y$ and the product of the marginals. Intuitively, this makes sense since if $X$ and $Y$ are independent the product of the marginals and the joint coincide meaning no difference and hence mutual information is equal to zero. A large difference means dependency and hence $Y$ tells more about $X$.

However, there could be other alternatives which can be just as good as the KL divergence. For instance several other measures of distances between probability distributions exist such as Jensen-Shannon divergence, Wasserstein/Earth mover distance, or Total variation distance. In this paper we define these alternatives for mutual information (Section \ref{sec:prelim}), consider different ways to estimate them from samples (Section \ref{MI_leak}), and evaluate the performance of the estimators (Section \ref{sec:Exp}).



\section{Preliminaries}\label{sec:prelim}
Let $\mathcal{X}$ be a probability space with $\sigma$-algebra $\mathcal{E}$ and measure $P$. We consider two random variables $X$ and $Y$ on this space, i.e. $X,Y\colon \mathcal{X}\to \mathbb{R}^d$. We equip the measurable space $\mathbb{R}^d$ with the Borel $\sigma$-algebra $\mathcal{B}$ and Lebesgue measure $\lambda$. The measure on $\mathbb{R}^d$ induced by $X$ given by $P(X^{-1}(B))$ for all $B\in \mathcal{B}$ is called the distribution $P_X$ (similarly for $Y$) and the probability density function (pdf) $p(x)$ if it exists is defined to be the function satisfying
\begin{equation}
    P(X^{-1}(B))=\int_{B}p(x)d\lambda(x)
\end{equation}
for all $B\in \mathcal{B}$, also known as the Radon-Nikodym derivative $p(x)=\frac{dP_X}{d\lambda}$. Hence, we remark that we will use the following two equivalent notations for integrals
\begin{equation}
    \int_{\mathcal{X}}dP_X=\int_{\mathbb{R}^d}p(x)d\lambda(x).
\end{equation}

\subsection{Definition of Mutual Information and Alternatives}\label{sec:definitions}
Now, we present the definitions of the different divergences/metrics and we start with the original definition of mutual information from the KL divergence.\footnote{We remark that the mutual information can also be defined using the differential entropy but this is equivalent to the KL-definition.}
\begin{defi}\label{kl_def}
    Let $P$ and $Q$ be two probability distributions, where $P$ is absolutely continuous with respect to $Q$. The Kullback-Leibler (KL) divergence between them is defined as
    \begin{align}
        D_{KL}(P \parallel Q)&=\int_{\mathcal{X}} \log\left(\frac{dP}{dQ}\right) dP\\&=\int_{\mathbb{R}^d}p(x)\log\left(\frac{p(x)}{q(x)}\right) d\lambda(x),
    \end{align}
    where $\frac{dP}{dQ}$ is the Radon-Nikodym derivative and the last equality holds if the pdf's $p(x)$ and $q(x)$ exists.
\end{defi}
\begin{defi}
    Let $X\colon \mathcal{X}\to \mathbb{R}^d$ and $Y\colon \mathcal{Y}\to \mathbb{R}^m$ be two continuous random variables with distributions $P_X$ and $P_Y$ respectively. If $P_{(X,Y)}$ is the joint distribution of $X$ and $Y$, then the mutual information between $X$ and $Y$ is given by
    \begin{equation}\label{eq:MI_from_KL}
        I_{KL}(X;Y)=D_{KL}(P_{(X,Y)}\parallel P_X\otimes P_Y).
    \end{equation}
\end{defi}
As we can see, the mutual information is a comparison (using the KL-divergence) of the joint distribution with the product of the marginals. However, as described in the introduction we can also make this comparison using other divergences/metrics on probability measures. We will introduce some of these below. We first look at some other divergences, which similar to the KL-divergence is a special class of divergences called $f$-divergences \cite{renyi1961measures}. An $f$-divergence between two probability distributions $P$ and $Q$, where $P$ is absolutely continuous with respect to $Q$, is defined as
\begin{equation}
    D_f(P\parallel Q)=\int_\mathcal{X} f\left(\frac{dP}{dQ}\right)dQ,
\end{equation}
and if both $P$ and $Q$ are absolutely continuous with respect to the Lebesgue measure $\lambda$ we have the densities satisfying $dP=p(x)d\lambda(x)$ and $dQ=q(x)d\lambda(x)$ implying that we can write
\begin{equation}
    D_f(P\parallel Q)=\int_{\mathbb{R}^d}f\left(\frac{p(x)}{q(x)}\right)q(x)d\lambda(x).
\end{equation}
With this notation the KL-divergence is an $f$-divergence with $f(t)=t\log(t)$.
\begin{defi}
     Let $P$ and $Q$ be two probability distributions. The Jensen-Shannon (JS) divergence of $P$ and $Q$ is an $f$-divergence with $f(t)=\frac{1}{2}\left((t+1)\log(\frac{2}{t+1})+t\log(t)\right)$. 
\end{defi}
From this, it can be deduced that
\begin{equation}\label{JS_as_KL}
\begin{aligned}
    D_{JS}(P \parallel Q)=& \frac{1}{2}D_{KL}(P\parallel \frac{P+Q}{2})+\frac{1}{2}D_{KL}(Q\parallel \frac{P+Q}{2})\\=&\frac{1}{2}\int_{\mathbb{R^d}} p(x)\log\left(\frac{2p(x)}{p(x)+q(x)}\right)\\ &+q(x)\log\left(\frac{2q(x)}{p(x)+q(x)}\right) d\lambda(x),
\end{aligned}
\end{equation}
and we remark, that the JS divergence is a symmetric version of the KL divergence.

\begin{defi}
     Let $P$ and $Q$ be two probability distributions. The Total variation (TV) divergence between them is an $f$-divergence with $f(t)=\frac{1}{2}|t-1|$.
\end{defi}
Again, this shows that
\begin{equation}\label{eq:d_tv}
\begin{aligned}
    D_{TV}(P\parallel Q)&= \frac{1}{2}\int_{\mathcal{X}}|dP-dQ|\\&=\frac{1}{2}\int_{\mathbb{R}^d}|p(x)-q(x)|d\lambda(x).
\end{aligned}
\end{equation}
Intuitively, it makes sense to compare how much the two densities differ from each other throughout the whole domain. But on the other hand, the total variation can sometimes be a too strong metric. 


As an alternative to the $f$-divergences we also consider a metric of probability measures, namely the Wasserstein distance. 
\begin{defi}\label{wass_dist_def}
    Let $P$ and $Q$ be two probability measures on a metric space and let $d(x,y)$ be a metric on this space. Then the $k$'th Wasserstein distance is
    \begin{equation}
        W_k(P,Q) = \inf_{\gamma\in\Gamma(P,Q)}\left(\int_{\mathcal{X}\times \mathcal{X}} d(x,y)^k d\gamma(x,y)\right)^{\frac{1}{k}}
    \end{equation}
    where $\Gamma(P,Q)$ is the set of all couplings of $P$ and $Q$, i.e. the set of measures having $P$ and $Q$ as the marginals. 
\end{defi}
One can see $\gamma$ as a transportation plan for transforming $P$ into $Q$. Therefore, $W_1(P,Q)$ is also known as the earth mover distance since it measures the cost of ``moving the mass'' from $P$ to $Q$. Since $W_1$ has the earth mover interpretation this will be the one we are focusing on the most in this paper.

In fact, both the $W_1$ and the TV distance can be described as an integral probability metric (IPM) \cite{muller1997integral}. 
\begin{defi}\label{IPM_def}
For two probability measures ${P}$ and $Q$ on a measurable space $\mathcal{X}$, IPM is given by
\begin{equation}
    \gamma(\mathcal{F} ,P,Q) = \sup_{f \in \mathcal{F}} { \int f(x) dP(x) - \int f(y) dQ(y) }
\end{equation}
where $\mathcal{F}$ is a space for measurable functions on $\mathcal{X}$.
\end{defi}
For different $\mathcal{F}$ we obtain different metrics. In fact, the Rubenstein-Kantorovich duality shows that the \\ Wasserstein-1 distance can be captured as an IPM
\begin{equation}
\begin{aligned}
     \gamma(\mathcal{L}_M ,P,Q) &= \sup_{f \in \mathcal{L}_M} { \int f(x) d{P}(x) - \int f(y) d{Q}(y) }\\
     &=W_{1}({P}, {Q})
\end{aligned}
\end{equation}
where $\mathcal{L}_M$ is the set of all functions that are 1-Lipschitz functions \cite{villani2009optimal}. Similarly, we can capture the TV by setting $\mathcal{F}_{TV} = \left\{f \mid\|f\|_{\infty} \leq \frac{1}{2}\right\}$ \footnote{We choose $\frac{1}{2}$ ($\beta-\alpha = 1$) as our bound so that we can have correspondence with our definition for total variation in eq. \eqref{eq:d_tv} referring to Theorem 2 in \cite{sriperumbudur2009integral}} where $\|f\|_{\infty} = \sup_{x \in \mathcal{X}} |f(x)| $. 

To measure information leakage we compare the joint distribution with the product of the marginals of two random variables $X$ and $Y$, like in \eqref{eq:MI_from_KL}. Since the other divergences/metrics can be used as alternatives for measuring the information leakage we will use a mutual information-like notation. Hence, we write $I_{JS}(X;Y)$, $I_{TV}(X;Y)$, and $I_{W_k}(X;Y)$ when we do the same comparison as in \eqref{eq:MI_from_KL} but with another divergence/metric.

\subsection{Relation Between Different Measures}
Many of the divergences are related to each other as for instance it is seen in \eqref{JS_as_KL} where the JS divergence can be defined from the KL divergence. But beside this there exists several bounds on the different divergences. We will present some of these here but refer the reader to \cite{SV16} for more theory about this. First, we relate the JS and the TV divergences. The following relations holds when the base of the logarithm used in the JS divergence is base $2$
\begin{equation}
    0\leq D_{JS}(P\parallel Q)\leq D_{TV}(P\parallel Q) \leq 1.
\end{equation}

We can also relate the TV divergence to the KL divergence with the following upper bounds on $D_{TV}$
\begin{equation}\label{eq:TV_KL_bound1}
    D_{TV}(P\parallel Q)\leq \sqrt{\frac{1}{2}D_{KL}(P\parallel Q)}
\end{equation}
\begin{equation}\label{eq:TV_KL_bound2}
    D_{TV}(P\parallel Q)\leq \sqrt{1-e^{-D_{KL}(P\parallel Q)}}.
\end{equation}

Furthermore, the Wasserstein-1 distance is related to the TV divergence by the following bounds 
\begin{equation}
    \begin{aligned}
    W_1(P,Q)&\geq D_{TV}(P\parallel Q)\cdot d_{min} \\
    W_1(P,Q)&\leq \mathrm{diam}(\mathcal{X}) D_{TV}(P\parallel Q),
    \end{aligned}
\end{equation}
where $d_{min}=\inf_{x\neq y}\{d(x,y)\} $ for $x,y\in \mathcal{X}$ and $\mathrm{diam}(\mathcal{X})=\sup_{x,y\in \mathcal{X}}\{d(x,y)\}$. We remark that the bounds on the Wasserstein distance is not very useful for instance if $\mathcal{X}=\mathbb{R}^d$ and we are using the Euclidean distance since $d_{min}=0$ and $\mathrm{diam}(\mathcal{X})=\infty$ in this case. However, if $\mathcal{X}$ a bounded set then $\mathrm{diam}(\mathcal{X})<\infty$. Furthermore, we remark that if we use the discrete metric 
\begin{equation}
d(x,y)=\begin{cases}
1 \text{ if }x\neq y \\ 
0\text{ if }x= y  \end{cases}
\end{equation}
then the inequalities actually show that the total variation and the Wasserstein distance are equivalent. Hence, we can consider the total variation as a special case of the Wasserstein-1 distance where we use the discrete metric in the Wasserstein. However, we will stick to the Euclidean norm for the Wasserstein distance in this paper. 

We also mention, that increasing the $k$ in the Wasserstein distance will increase the distance, i.e. $W_{k_1}(P,Q)\leq W_{k_2}(P,Q)$ when $k_1\leq k_2$ \cite{villani2009optimal}.

\section{Mutual information as leakage measure in multiparty computation}\label{MI_leak}
In multiparty computation $n$ parties would like to compute $f(x_1,\ldots,x_n)$ where $x_i$ is held by the $i$'th party. The computation needs to be secure even in the presence of an adversary corrupting a number of the parties. This means for instance that the adversary is not allowed to learn more than it will learn from the input of the corrupted parties and the output $f(x_1,\ldots,x_n)$. The security is usually proven through a simulation proof but the privacy requirement can also be stated via mutual information in the following way
\begin{equation}\label{MPC_sec}
    I(X_i;\mathrm{View}_\mathcal{A})=I(X_i;f(X_1,\ldots,X_n),\{X_j\}_{j\in \mathcal{A}})
\end{equation}
where $\mathrm{View}_\mathcal{A}$ is everything the adversary $\mathcal{A}$ sees through the algorithm. Some privacy-preserving algorithms do not guarantee the equality in \eqref{MPC_sec} but instead they ensure that the leakage is small meaning that $I(X_i;\mathrm{View}_\mathcal{A})$ is not much higher than $I(X_i;f(X_1,\ldots,X_n),\{X_j\}_{j\in \mathcal{A}})$. Hence, it is interesting to be able to compute $I(X_i;\mathrm{View}_\mathcal{A})$ but the density of $\mathrm{View}_\mathcal{A}$ might not always be known which makes it difficult for computation. Hence, approximating this mutual information is interesting from a multiparty computation perspective. 

Thus, we present different ways to approximate the different metrics/divergences defined in Section \ref{sec:definitions} on probability distributions $P$ and $Q$ in the following sections. The approximations are based on samples $\mathbf{X}=\{x_i \}_{i=1}^{N}$ from a random variable $X$ having distribution $P$ and density $p(x)$ and samples $\mathbf{Y}=\{y_i \}_{i=1}^{N}$ from a random variable $Y$ having distribution $Q$ and density $q(y)$. Often in practice $p$ and $q$ are not known so we compute approximations of them.

\subsection{Leakage Estimation via Histograms}\label{leak_histo}
In this section we give a general way to approximate the pdf's using histograms. We build up histograms by splitting the domain into $K$ bins and count the number of instances in each bins. I.e. the domain equals $\bigcup_{i=1}^K B_i$ where $B_i\cap B_j=\emptyset$ when $i\neq j$. This gives rise to an approximate pdf of $p$, where we let $n_{p,i}$ be the number of instances in bin $B_i$ from the samples of $X$;
\begin{equation}
    \hat{p}(x)= \sum_{i=1}^K\mathds{1}_{x\in B_i}\frac{n_{p,i}}{N\cdot \lambda(B_i)},
\end{equation}
where $\mathds{1}_{x\in B_i}$ is the indicator function for $x$ in $B_i$. We remark that the $B_i$'s are disjoint and hence for each $x$ exactly one of the terms in the sum is nonzero. Hence, $\hat{p}(x)=\frac{n_{p,i}}{N\cdot \lambda(B_i)}$ if $x\in B_i$. Since all the divergences we look at are $f$-divergences, they can be described as an integral with respect to the Lebesgue measure of some function of the pdf's, i.e. $f(p(x),q(x))$. Hence, we can split the interval up in a sum of intervals where we integrate over a constant. It means that
\begin{equation}
\begin{aligned}
    &\int_{\mathbb{R}^d} f(p(x),q(x)) d\lambda(x) =\sum_{i=1}^K\int_{B_i} f(p(x),q(x)) d\lambda(x) \\
    &\approx \sum_{i=1}^K\int_{B_i} f(\hat{p}(x),\hat{q}(x)) d\lambda(x) = \sum_{i=1}^K \mathrm{\lambda}(B_i) c_i,
\end{aligned}
\end{equation}
where the last equality follows from the fact that $\hat{p}(x)$ and $\hat{q}(x)$ is constant inside bin $B_i$ and hence $f(\hat{p}(x),\hat{q}(x))=c_i$ is constant inside this bin. This implies the following approximations of the different divergences. 
\begin{align}
\begin{aligned}
    D_{KL,hist}(P\parallel Q)=&\sum_{i=1}^K \frac{n_{p,i}}{N}\log\left(\frac{n_{p,i}}{n_{q,i}}\right)\\
    D_{TV,hist}(P\parallel Q)=&\frac{1}{2}\sum_{i=1}^K \left|\frac{n_{p,i}}{N}-\frac{n_{q,i}}{N}\right|\\
    D_{JS,hist}(P\parallel Q)=&\frac{1}{2}\left(\sum_{i=1}^K \frac{n_{p,i}}{N}\log\left(\frac{2n_{p,i}}{n_{q,i}+n_{p,i}}\right)\right. \\&+\left.\frac{n_{q,i}}{N}\log\left(\frac{2n_{q,i}}{n_{q,i}+n_{p,i}}\right)\right).
\end{aligned}
\end{align}
We can also use \eqref{LP_dist_general} to approximate $W_1$ using the histogram method. We describe this through optimal transport below.


\subsection{Wasserstein Distance Via Optimal Transport}\label{opt_tr_wass}

Optimal transport is often formulated in a discrete setting, so we start by considering the optimal transport between two discrete distributions. Hence, we consider a situation where we have $p(x)$ and $q(y)$ and they can be described by two vectors $\mathbf{p}$ and $\mathbf{q}$ where the entries are the probabilities for the different outcomes and hence the entries in $\mathbf{p}$ (and $\mathbf{q}$) sum to $1$. Optimal transport describes the cost (how far the mass needs to be moved and how much mass) of transporting $p(x)$ to $q(y)$.

The solution to the optimal transportation problem between $\mathbf{p}$ and $\mathbf{q}$ (and hence the distributions) is nothing but a matrix $[M_{ij}]_{i,j=1}^{N} \in \mathbb{R}_{+}^{N \times N}$ where the element $M_{ij}$ represents the amount of mass transported from $p_i$ to $q_j$. In order to find the optimal transportation plan $M$ using the cost function $C$, one can form linear program as
\begin{equation}\label{LP_dist_general}
    \begin{aligned}
d(x, y) =& \underset{M \geq 0}{\min} \langle C,M \rangle \\
& \text { subject to }
 M \mathbf{1}=\mathbf{p}, \\
& \qquad \qquad \quad M^{T} \mathbf{1}=\mathbf{q},
    \end{aligned}
\end{equation}
where $\mathbf{1}$ is the vector of all ones. The optimal transportation plan between $\mathbf{p}$ and $\mathbf{q}$ is obtained after solving the linear program problem in (\ref{LP_dist_general})
\cite{haasler2021control}.

Now we look at the continuous case. We still have $p(x)$ and $q(y)$ but they are now continuous functions. The equivalence of minimizing $\langle C,M \rangle$ with respect to $M$ in the discrete setting is to take the infimum with respect to all possible $m(x,y)$ of 
\begin{equation*}
    \int_{\mathbb{R}^d\times\mathbb{R}^d}c(x,y)m(x,y)d\lambda(x,y)
\end{equation*}
satisfying $\int_{\mathbb{R}^d}m(x,y)\d\lambda(x)=q(y)$ and $\int_{\mathbb{R}^d}m(x,y)\d\lambda(y)=p(x)$. But with $c(x,y)$ being a metric (and in our case the Euclidean distance) this is nothing else than the $W_1$ distance from definition \ref{wass_dist_def}. However, this problem can be hard to solve. But approximating $p(x)$ and $q(x)$ by histograms from samples we can approximate the Wasserstein distance by solving \eqref{LP_dist_general} setting the entries in $\mathbf{p}$ equal $\frac{n_{p,i}}{N}$ and similarly for $\mathbf{q}$. In this case we need to define the distance matrix $C$ but a natural way to do so, is to compute the distance between the centers of the bins. 



The optimal transportation is very useful when we have a relatively small number of bins. However, when we have a large amount of bins, solving the linear program problem can be computationally heavy. To address this issue, it is suggested to apply the Sinkhorn distance. The optimization problem can be converted into a Sinkhorn distance between two probability vectors $\mathbf{p}$ and $\mathbf{q}$ by introducing a Lagrange multiplier for the entropy constraint as 
\begin{equation}\label{eq:Sinkhorn_dist}
    \begin{aligned}
d_\lambda(x, y) =& \underset{M \geq 0}{\min} \langle C,M \rangle -\frac{1}{\lambda}h(M)\\
& \text { subject to }
 M \mathbf{1}=\mathbf{p}, \\
& \qquad \qquad \quad M^{T} \mathbf{1}=\mathbf{q},
    \end{aligned}
\end{equation}
where $\lambda \geq 0$ is a tuning parameter that scales the entropy constraint $h(M)=-\sum_{i,j=1}^N m_{ij}\log(m_{ij})$ (\cite{cuturi2013sinkhorn}, equation (2)). By optimizing $d_{\lambda}(x,y)$, implies an upper bound on the Wasserstein-1 distance which computationally should be easier to compute. Furthermore, we mention that if $\lambda$ is large, the Sinkhorn distance would be a good approximation of the Wasserstein distance. We used Algorithm (1) in \cite{cuturi2013sinkhorn} and please refer to here for a detailed description of the Sinkhorn distance.

\subsection{A KL-estimator From Samples}\label{knn_apprx}
In this section we describe a KL-divergence estimator given a set of samples from two distributions. The estimator is presented in \cite{KL_estimator}. The estimator is computed by approximating $p(x)$ and $q(y)$ around $x_i$ by looking at the $k$-th nearest neighbor to $x_i$. They show that even though the approximations of $p(x)$ and $q(y)$ do not necessarily converge to $p(x)$ and $q(y)$ the estimator will converge to the true KL-divergence when increasing the sample size. The estimator is given by
\begin{equation}
    D_{KL,k-nn}(P\parallel Q)=\frac{d}{n}\sum_{i=1}^{N}\log\left(\frac{s_k(x_i)}{r_k(x_i)}\right)+\log\left(\frac{N}{N-1}\right),
\end{equation}
where $r_k(x_i)$ is the Euclidean distance from $x_i$ to the $k$-the nearest neighbor in $\mathbf{X}\setminus \{x_i\}$, and $s_k(x_i)$ is the Euclidean distance to the $k$-the nearest neighbor in $\mathbf{Y}$.

\subsection{Leakage Estimation via Kernel Mean Embedding}\label{kern_mmd}

The use of kernel functions range widely in classical machine learning topics such as support vector machines, the principal component analysis and perceptron for deploying inner product $\langle x\,,x^{'} \rangle$ of two data instances $x,x^{'} \in \mathcal{X}$ which measures the distance between those instances. However, linear functions applied to the inner product sometimes fail when generalizing the distance measure.

In order to overcome this issue, one can apply a "kernel trick" and make the distance measure accurate enough by replacing the inner product with a possible non-linear mapping. Kernel methods rely on kernel functions and can be defined as an inner product of a mapping function which transforms data instances into a higher dimensional feature space as 
\begin{equation}
    k(x,x^{'})=\langle \phi(x) \,,\phi(x^{'}) \rangle _{\mathscr{H}}
\end{equation}
where $k(\cdot , \cdot)$ is the kernel function, $\phi(\cdot)$ is the mapping function for data instances as $\{ \phi : \mathcal{X} \rightarrow \mathscr{H} , x \rightarrow \phi(x) \}$, and $\langle \phi(x) \,,\phi(x^{'}) \rangle _{\mathscr{H}}$ is the inner product in the reproducing kernel Hilbert space (RKHS) $\mathscr{H}$. By introducing this trick, which depends on substituting the $\langle x \,,x^{'} \rangle$ with $\langle \phi(x) \,,\phi(x^{'}) \rangle _{\mathscr{H}}$, it is possible to apply the inner product in a higher dimension, and measure the similarities between $x \text{ and } x^{'}$ \cite{muandet2016kernel}. In this way, we do not need to explicitly construct $\phi(x)$ and not need to know $\mathscr{H}$ specifically. It will be sufficient to use positive definite kernels in $\mathscr{H}$ for the benefits. As a such kernel, we used the Gaussian kernel $k\left({x}, {x}^{\prime}\right)=\exp (-\frac{\|{x}-{x}^{\prime}\|_{2}^{2}}{2 \sigma^{2}})$ with $\sigma = \sqrt{1/2}$ for this paper. 

When it comes to defining a metric for probability distributions, it is useful to consider the IPM setting in definition \ref{IPM_def}, where we could change the function class $\mathcal{F}$ to specify different metrics. 


Hence, we define yet another metric from IPM, namely the maximum mean discrepancy (MMD) (with the help of kernel function eventually) using $\mathcal{F} := \left\{ f \mid \|f\|_{\mathscr{H}} \leq 1 \right\}$ to get the functions from a unit ball in $\mathscr{H}$. Let $X, X', Y, \text{ and } Y'$ be independent representations as $X,X' \sim {P}$ and $Y,Y' \sim {Q}$ and set $\mu_{{P}}=\mathbb{E}_{X}[k(\cdot, X)]$ and similarly for $\mu_Q$. For a more in depth walk through of this, we refer the reader to \cite{muandet2016kernel}. In here, it is also described that for a function $f$, the kernel function can be seen as the evaluation of $f$ in the following way $f(x)=\langle f,k(\cdot,x)\rangle$. From this one can define MMD as the distance between mean embedding in $\mathscr{H}$ as 
\begin{equation}\label{eq:mmd}
    \begin{aligned}
        \text{MMD}[\mathscr{H},{P},{Q}] &= \sup_{\| f \|_{\mathscr{H}} \leq 1} { \int f(x) d{P}(x) - \int f(y) d{Q}(y) } \\ &= \sup_{\| f \|_{\mathscr{H}} \leq 1} \mathbb{E}_{P}[f(X)]-\mathbb{E}_{Q}[f(Y)]\\ &= \sup_{\| f \|_{\mathscr{H}} \leq 1} \mathbb{E}_{P}[\langle f,k(\cdot,X)\rangle]-\mathbb{E}_{Q}[\langle f,k(\cdot,Y)\rangle]\\ &= \sup_{\| f \|_{\mathscr{H}} \leq 1} \langle f , \mu_{{P}} - \mu_{{Q}} \rangle_{\mathscr{H}} \\ &= \| \mu_{{P}} - \mu_{{Q}} \|_{\mathscr{H}}
    \end{aligned}
\end{equation}
with the use of the reproducing property of $\mathscr{H}$ and the linearity of the inner product and expectation. The $\mu_{{P}}$ in the equation (\ref{eq:mmd}) can be used as $\left\|\mu_{{P}}\right\|_{\mathscr{H}}^{2}=\langle\mathbb{E}_{X}[k(\cdot, X)], \mathbb{E}_{X'}[k(\cdot, X')]\rangle_{\mathscr{H}}=\mathbb{E}_{X, X'}[k(X, X')]$. Then, the use of kernel functions become practical to calculate the MMD metric using the expression
\begin{equation}
\begin{aligned}
    \text{MMD}^{2}[\mathscr{H},{P},{Q}] &= \mathbb{E}_{P, P}[k(X, X')]+\mathbb{E}_{Q, Q}[k(Y, Y')] \\ &-2 \mathbb{E}_{P, Q}[k(X, Y)].
\end{aligned}
\end{equation}
One can form empirical MMD using the i.i.d. samples from $\mathbf{X} = \{x_i \}_{i=1}^{N}$ and $\mathbf{Y} = \{ y_i \}_{i=1}^{N}$ with respect to the kernel function as 
\begin{equation}
\begin{aligned}
    \widehat{\mathrm{MMD}_{u}^{2}}[\mathscr{H}, \mathbf{X}, \mathbf{Y}] &= \frac{1}{N(N-1)}\sum_{i = 1}^{N} \sum_{j \neq i}^{N} k({x_i},{x_j}) \\ &+ \frac{1}{N(N-1)}\sum_{i = 1}^{N} \sum_{j \neq i}^{N} k({y_i},{y_j}) \\ &- \frac{2}{N^{2}}\sum_{i = 1}^{N} \sum_{j=1}^{N} k({x_i},{y_j})
\end{aligned}
\end{equation}
which is the unbiased estimator of the MMD \cite{borgwardt2006integrating}. This estimation of MMD is informative in a sense that it provides a lower bound for $W_{k1}(P, Q) \text{ and } \gamma(\mathcal{F}_{TV} ,P, Q)$ due to the inequalities $\text{MMD}[\mathscr{H},P,Q] \leq W_{k1}(P, Q) \leq \sqrt{\text{MMD}[\mathscr{H},P,Q]^2 + 4C}$ and $\text{MMD}[\mathscr{H},P,Q] \leq \sqrt{C} \gamma(\mathcal{F}_{TV} ,P, Q)$ satisfying $ \\ \sup_{{x} \in X} k({x},{x}) \leq C \leq \infty $ for a constant $C$, and $W_{k1}$ is the kernel-based $W_1$ by defining $d(x, y)=\|k(\cdot, x)-k(\cdot, y)\|_{\mathscr{H}}$ \cite{sriperumbudur2010hilbert}. While estimation of MMD only provides a lower bound for total variation, it provides both lower and upper bound for Wasserstein distance. Convergence of MMD is another advantage of using it as a measure. According to \cite{muandet2016kernel}, MMD appears to be converging with the rate of $\sqrt{1/N}$. Dimension of $d$ does not effect the convergence apart from a constant $c_d$ that depending on $d$ as $\left|\widehat{\mathrm{MMD}_{u}^{2}}[\mathscr{H}, \mathbf{X}, \mathbf{Y}]-\operatorname{MMD}^{2}[\mathscr{H}, {P}, {Q}]\right| \leq c_{d}\sqrt{1/N}$.

\section{Experiment and Results}\label{sec:Exp}
Since the motivation is especially leakage of information in a distributed computation such as a multiparty computation protocol we evaluate the metrics in such setups. We take our inspiration from \cite{Katrine_Shamir} where secret sharing is defined over the real numbers. In contrast to the traditional finite field multiparty computation a share might reveal some (but limited) information. Hence, we will evaluate the leakage of a share and a small multiparty computation using this concept.

\subsection{A Secret and Its Share Scenario}\label{ex_41}
First, we consider the leakage of information of a normal distributed $X \sim \mathcal{N}(\mu_x,\sigma_x^2)$ from an obfuscated version of $X$, namely $X - R$ where $R\sim \mathcal{N}(\mu_r,\sigma_r^2)$. This can also be seen as a share of $X$ in the real number secret sharing scheme from \cite{Katrine_Shamir}. The joint distribution of $(X,X-R)$ is
\begin{equation}\label{eq:dist_4.1_joint}
    \mathcal{N}\left(\begin{bmatrix}\mu_x\\ \mu_x-\mu_r \end{bmatrix},\begin{bmatrix}\sigma_x^2 & \sigma_x^2\\ \sigma_x^2&\sigma_x^2+\sigma_r^2 \end{bmatrix}\right),
\end{equation}
and the product of the marginals is
\begin{equation}\label{eq:dist_4.1_marg}
    \mathcal{N}\left(\begin{bmatrix}\mu_x\\ \mu_x-\mu_r \end{bmatrix},\begin{bmatrix}\sigma_x^2 & 0\\ 0&\sigma_x^2+\sigma_r^2 \end{bmatrix}\right),
\end{equation}
and hence we can actually determine the KL-divergence explicitly in this case. In our experiments, we assume that both $X$ and $R$ has $0$ mean so for simplicity we do the same here. In this situation we have
\begin{equation}
    I_{KL}(X;X-R)=\frac{1}{2}\ln\left(1+\frac{\sigma_x^2}{\sigma_r^2}\right).
\end{equation}
In the experiments we set $\sigma_x^2=1$ and $\sigma_r^2=10$ implying that
\begin{equation}\label{eq:kl_bound}
    I_{KL}(X;X-R)=\frac{1}{2}\ln\left(1.1\right)=0.048.
\end{equation}
The upper bound on TV in \eqref{eq:TV_KL_bound1} is less than the bound in \eqref{eq:TV_KL_bound2} in this case, and it implies that
\begin{equation}\label{eq:js_tv_bound}
    I_{JS}(X;X-R)\leq I_{TV}(X;X-R)\leq 0.154.
\end{equation}

However, we can also use the fact that the JS-divergence can be defined from the KL-divergence as in \eqref{JS_as_KL}. We have
\begin{equation}
\begin{aligned}
    &I_{JS}(X;X-R)=\\&\frac{1}{2}D_{KL}\left(P_{(X,X-R)}||\frac{P_X\otimes P_{X-R}+P_{X,X-R}}{2}\right)\\&+\frac{1}{2}D_{KL}\left(P_{X}\otimes P_{X-R}||\frac{P_X\otimes P_{X-R}+P_{X,X-R}}{2}\right)
\end{aligned}
\end{equation}
Both of these KL-divergences are between a Gaussian and a mixture of two Gaussians with coefficients $\frac{1}{2}$. In \cite{KL_bound_GMM} they give upper bounds on KL-divergence between mixture of Gaussians and from this we can derive
\begin{equation}
\begin{aligned}
    &I_{JS}(X;X-R)\leq\\ &-\frac{1}{2}\log\left(\frac{1}{2}(1+e^{-D_{KL}(P_{(X,X-R)})||P_X\otimes P_{X-R})})\right)\\
    &-\frac{1}{2}\log\left(\frac{1}{2}(1+e^{-D_{KL}(P_X\otimes P_{X-R}||P_{(X,X-R)})})\right).
\end{aligned}
\end{equation}
Here we only have KL-divergences between Gaussians which we have closed form solutions for. Plugging in the distributions from \eqref{eq:dist_4.1_joint} and \eqref{eq:dist_4.1_marg} with zero mean and $\sigma_x^2=1$, $\sigma_r^2=10$, we obtain that 
\begin{equation}\label{eq:js_bound}
    I_{JS}(X;X-R)\leq 0.0356
\end{equation}

Furthermore, there exists a closed form expression for the Wasserstein-2 distance with the Euclidean norm when the two distributions are normal. Let $\Sigma_1$ and $\Sigma_2$ be the covariance matrices of the two distributions (and assume 0-mean) then the Wasserstein-2 distance is
\begin{equation}
    \sqrt{\mathrm{trace}\left(\Sigma_1+\Sigma_2-2\left(\Sigma_2^{\frac{1}{2}}\Sigma_1\Sigma_2^{\frac{1}{2}}\right)^\frac{1}{2}\right)}.
\end{equation}
This implies that
\begin{equation}
    I_{W_1}(X;X-R)\leq W_2(P_{(X,X-R)},P_X\otimes P_{(X-R)})=0.292
\end{equation}
when we use the covariance matrices from \eqref{eq:dist_4.1_joint} and \eqref{eq:dist_4.1_marg}.

\subsection{Multiplication Using Three Parties Scenario}\label{ex_42}
In this scenario we consider a situation with three parties where two of them having a value $s$ and $t$ respectively and the parties wants to learn the product $st$. We treat $s$ and $t$ as outcomes of random variables $S$ and $T$ and we use the Shamir secret sharing scheme from \cite{Katrine_Shamir} with privacy threshold $1$ and evaluation points $p_1=-1$, $p_2=1$, and $p_3=2$. In this situation the shares can be constructed by evaluating $f_s(x)=s+(r_s-s)x$ at $p_i$ and hence the shares of $s$ is $(2s-r_s,r_s,-s+2r_s)$. Notice that having two shares leaks everything about $s$. The algorithm goes on like this. The party having $s$, secret shares $s$ by sending $f_s(p_i)$ to the $i$'th party. Similarly, is $t$ shared by sending $f_t(p_i)$ to the $i$'th party. Now, the $i$'th party computes $f_s(p_i)f_t(p_i)$ and sends this value to the other two parties. Since this is a evaluation of a degree-2 polynomial and each party has three evaluations they are able to determine the polynomial having constant term $st$. We assume that the first party is not having an input and we want to evaluate how much he learns about $s$ from following this distributed algorithm. I.e. we want to evaluate
\begin{equation}
    I(S;2S-R_s,2T-R_t,R_tR_s,(-S+2R_s)(-T+2R_t)).
\end{equation}
In the experiments we assume that $S$ and $T$ are following a $\mathcal{N}(0,1)$ distribution and $R_s$ and $R_t$ are normal as well with $0$ mean and variance $\sigma_r^2=10$.
\subsection{Experimental Setup}

For both scenarios explained in Section \ref{ex_41} and Section \ref{ex_42}, we will consider the convergences for all divergences and metrics with respect to the number of samples and the number of bins we used. We will also report and comment on the run time for each approximation and computation. However, we were unable to compute the Wasserstein and Sinkhorn distance due to memory problems for the scenario in Section \ref{ex_42} which indicates that the two methods are impractical for measuring information leakage. Furthermore, the $k$-nearest neighbor approach does not seem to converge for the amount of samples we were able to evaluate. Therefore, we only show the convergence for the histogram-based divergences in this case.

First of all, we set the number of samples $ N $ for all experiments to be in the span of $N = [100, 500, 10^{3}, 2 \times 10^{3}, 3 \times 10^{3}, 5 \times 10^{3}, 10^{4}, 10^{5},10^{6},10^{7},10^{8}]$ for histogram based approximations (while we set the number of bins as $24$ in each dimension) and $N = [100, 500, 10^{3}, 2 \times 10^{3}, 3 \times 10^{3}, 4 \times 10^{3}, 5 \times 10^{3}, 7 \times 10^{3}, 10^{4}, 2 \times 10^{4}]$ for the sample based approximations. Span of the number of bins (in each dimension) is $[8, 10, 12, 16, 20, 24, 28, 30]$ while we set the number of samples as $N = 10^7$ for scenario \ref{ex_41}. We set the $\lambda$ parameter in Sinkhorn approximation as $\lambda = 700$ for scenario \ref{ex_41}.

Since we deal with histograms, some bins will be empty when we represent distributions with them. This can cause problems when we try to compute the KL divergence referring to the definition \ref{kl_def}. For the histogram of $q(x)$, empty bins will produce $\log(0)$ and this will cause an error in the implementation. To overcome this issue, we attain a small number ($10^{-8}$) that is close to $0$ whenever we have an empty bin for $q(x)$.

\subsection{Results}

We evaluate and interpret the results for our experiments.

\begin{figure}[h!]
\begin{center}
\includegraphics[width=8.4cm]{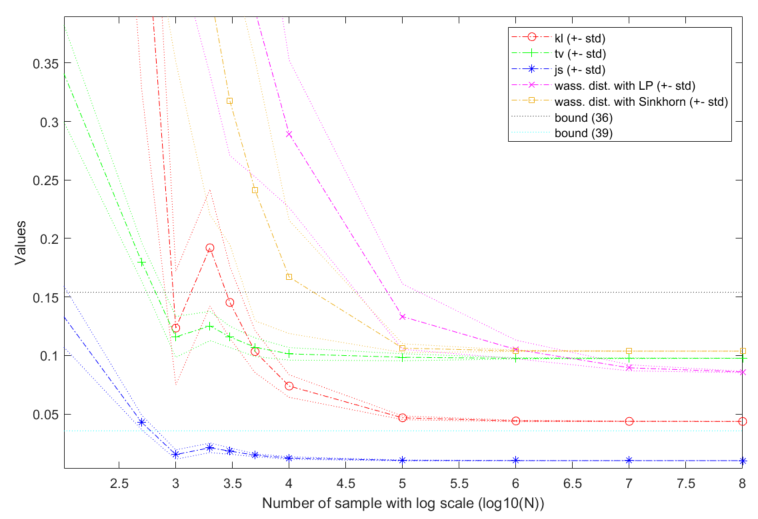} 
\caption{\scriptsize Average values of metrics for the scenario in 4.1 with respect to logarithm of the number of samples. Dotted lines represent the standard deviation for each metric. Red line is KL, green is TV and blue is JS divergences from Section \ref{leak_histo}. Dotted lines represent the standard deviations. Dotted black line is from bound (\ref{eq:js_tv_bound}). Magenta line is the histogram based $W_1$ calculation and cyan line is the Sinkhorn's estimation for $W_1$.}
\label{fig:div_4_1}
\end{center}
\end{figure}

In Figure \ref{fig:div_4_1}, we observe for the scenario described in Section \ref{ex_41} that we have a nice convergence for the histogram based estimations from Section \ref{leak_histo} of the KL, TV and JS divergences when sufficient amount of samples are used. After $10^5$ samples, divergences started to converge smoothly as well as their standard deviation narrows down. Furthermore, the KL divergence seems to converge to the right value in \eqref{eq:kl_bound}. On the other hand, the histogram based Wasserstein distance calculation via LP started to converge when we used a $10^7$ number of samples while the Sinkhorn estimation converged using $10^5$ samples. We also see that the bounds in (\ref{eq:js_tv_bound}) and (\ref{eq:js_bound}) hold after convergence. In Table \ref{tab:div_rt_4_1}, we do not observe significant run time change when the number of samples used is increased from $100$ to $10^7$. Noticeable change is seen when we used $10^8$ samples which implies around an increase in run time by a factor of 100 for the divergences and a doubling in runtime for the LP histogram based Wasserstein distance calculation. We also see that Sinkhorn approximation is much faster when compared with LP calculation.

\begin{table}[h!]
\centering
\resizebox{\columnwidth}{!}{%
\begin{tabular}{|l|l|l|l|l|l|l|l|l|l|l|l|}
\hline
 &
  100 &
  500 &
  $10^3$ &
  $2 \times 10^{3}$ &
  $3 \times 10^{3}$ &
  $5 \times 10^{3}$ &
  $10^{4}$ &
  $10^{5}$ &
  $10^{6}$ &
  $10^{7}$ &
  $10^{8}$ \\ \hline
\begin{tabular}[c]{@{}l@{}}kl  with \\ histogram\end{tabular} &
  $4.45 \times 10^{-4}$ &
  $4.45 \times 10^{-4}$ &
  $1.92 \times 10^{-4}$ &
  $1.91 \times 10^{-4}$ &
  $2.15 \times 10^{-4}$ &
  $5.14 \times 10^{-5}$ &
  $2.48 \times 10^{-4}$ &
  $2.47 \times 10^{-4}$ &
  $2.50 \times 10^{-4}$ &
  $2.77 \times 10^{-4}$ &
  $0.18 \times 10^{-1}$ \\ \hline
\begin{tabular}[c]{@{}l@{}}tv with\\ histogram\end{tabular} &
  $2.32 \times 10^{-4}$ &
  $2.59 \times 10^{-4}$ &
  $9.0 \times 10^{-5}$ &
  $1.12 \times 10^{-4}$ &
  $1.23 \times 10^{-4}$ &
  $4.93 \times 10^{-6}$ &
  $1.29 \times 10^{-4}$ &
  $1.46 \times 10^{-4}$ &
  $1.41 \times 10^{-4}$ &
  $1.57 \times 10^{-4}$ &
  $0.03 \times 10^{-1}$ \\ \hline
\begin{tabular}[c]{@{}l@{}}js with \\ histogram\end{tabular} &
  $4.45 \times 10^{-4}$ &
  $3.63 \times 10^{-4}$ &
  $1.50 \times 10^{-4}$ &
  $1.69 \times 10^{-4}$ &
  $2.00 \times 10^{-4}$ &
  $4.93 \times 10^{-5}$ &
  $1.99 \times 10^{-4}$ &
  $2.04 \times 10^{-4}$ &
  $2.06 \times 10^{-4}$ &
  $2.05 \times 10^{-4}$ &
  $0.04 \times 10^{-1}$ \\ \hline
\begin{tabular}[c]{@{}l@{}}wass. dist\\ with LP\end{tabular} &
  2.99 &
  2.72 &
  2.63 &
  2.66 &
  2.11 &
  1.89 &
  2.46 &
  2.60 &
  1.96 &
  2.04 &
  4.26 \\ \hline
\begin{tabular}[c]{@{}l@{}}wass. dist\\ with Sinkhorn\end{tabular} &
  0.016 &
  0.019 &
  0.017 &
  0.021 &
  0.017 &
  0.018 &
  0.017 &
  0.015 &
  0.010 &
  0.015 &
  0.138 \\ \hline
\end{tabular}%
}
\caption{\scriptsize Average run times for the scenario in 4.1 in seconds. Rows represent the approximations in \ref{leak_histo} and \ref{opt_tr_wass} and columns represent the number of samples used.}
\label{tab:div_rt_4_1}
\end{table}

\begin{figure}[h!]
\begin{center}
\includegraphics[width=8.4cm]{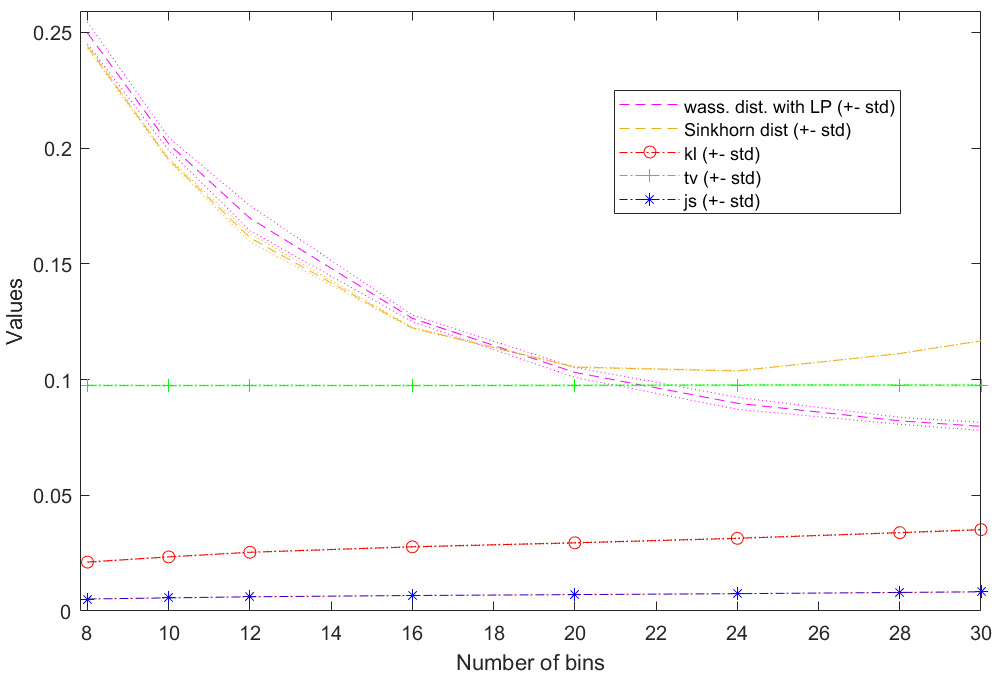} 
\caption{\scriptsize Average values of metrics for the scenario in 4.1 with respect to number of bins. Dotted lines represent the standard deviation for each metric. Red line is KL, green is TV and blue is JS divergences from Section \ref{leak_histo}. Dotted black line is from bound (\ref{eq:js_tv_bound}) and dotted cyan line is from bound (\ref{eq:js_bound}). Magenta line is the histogram based $W_1$ calculation and cyan line is the Sinkhorn's estimation for $W_1$.} 
\label{fig:lp_4_1}
\end{center}
\end{figure}

In Figure \ref{fig:lp_4_1}, we evaluate for the scenario \ref{ex_41} that histogram based $W_1$ via LP started to converge when using 30 bins in both dimensions for the histograms (in total 900 bins). Also Sinkhorn behaves like an upper bound for $W_1$ at the convergence level. The change in the number of bins does not seem to affect the behavior of TV and JS divergences. KL divergence is slightly increased but still around the expected value. The bounds in (\ref{eq:js_tv_bound}) and (\ref{eq:js_bound}) also hold after convergence. Table \ref{tab:lp_rt_4_1} represents the advantage of Sinkhorn approximation over LP calculation for $W_1$ in terms of the speed. For divergences, it seems that the choice of the number of bins has a minor effect on run time.


\begin{table}[h!]
\centering
\resizebox{\columnwidth}{!}{%
\begin{tabular}{|l|l|l|l|l|l|l|l|l|}
\hline
 & 8 & 10 & 12 & 16 & 20 & 24 & 28 & 30 \\ \hline
\begin{tabular}[c]{@{}l@{}}wass. dist.\\ with LP\end{tabular} & $0.0135$ & $0.0313$ & $0.0518$ & $0.2064$ & $0.7365$ & $1.9925$ & $4.8592$ & $12.0733$ \\ \hline
\begin{tabular}[c]{@{}l@{}}wass. dist.\\ with Sinkhorn\end{tabular} & $8.8232 \times 10^{-4}$ & $0.0012$ & $0.0013$ & $0.0026$ & $0.0078$ & $0.0151$ & $0.0357$ & $0.0993$ \\ \hline
\begin{tabular}[c]{@{}l@{}}kl with\\ histograms\end{tabular} & $2.2967 \times 10^{-4}$ & $1.9046 \times 10^{-4}$ & $1.8846 \times 10^{-4}$ & $0.0011$ & $2.2896 \times 10^{-4}$ & $2.4345 \times 10^{-4}$ & $3.5389 \times 10^{-4}$ & $0.0069$ \\ \hline
\begin{tabular}[c]{@{}l@{}}tv with \\ histograms\end{tabular} & $1.2121 \times 10^{-4}$ & $0.8717 \times 10^{-4}$ & $0.8518 \times 10^{-4}$ & $1.3931 \times 10^{-4}$ & $1.0123 \times 10^{-4}$ & $1.1391 \times 10^{-4}$ & $1.1795 \times 10^{-4}$ & $0.0010$ \\ \hline
\begin{tabular}[c]{@{}l@{}}js with\\ histograms\end{tabular} & $1.6072 \times 10^{-4}$ & $2.0740 \times 10^{-4}$ & $1.2036 \times 10^{-4}$ & $1.4439 \times 10^{-4}$ & $1.4119 \times 10^{-4}$ & $1.6904 \times 10^{-4}$ & $2.8426 \times 10^{-4}$ & $0.0012$ \\ \hline
\end{tabular}%
}
\caption{\scriptsize Average run times for the scenario in 4.1 in seconds. Rows represent the approximations in \ref{leak_histo} and \ref{opt_tr_wass} and columns represent the number of bins used.}
\label{tab:lp_rt_4_1}
\end{table}

\begin{figure}[h!]
\begin{center}
\includegraphics[width=8.4cm]{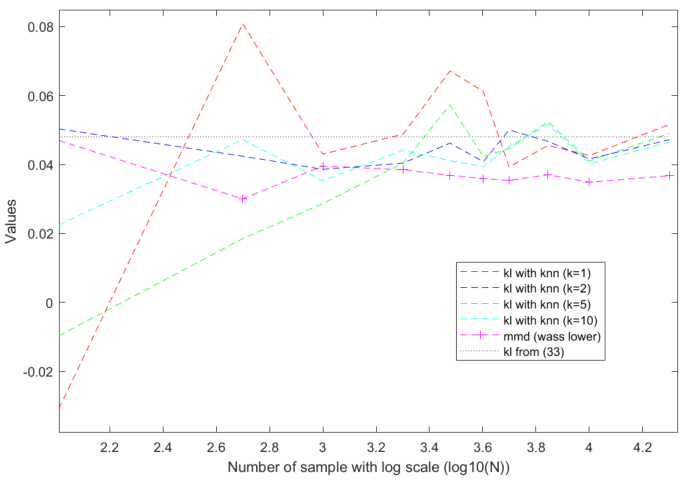} 
\caption{\scriptsize Average values of kernel and knn based metrics for the scenario in 4.1 with respect to logarithm of the number of samples. Red, blue, green and cyan lines are KL divergences using knn approximation in \ref{knn_apprx} for $k = 1,2,5,10$ respectively. Magenta line mmd with Gaussian kernel from \ref{kern_mmd}. Dotted black line is for the kl value in (\ref{eq:kl_bound}) }
\label{fig:knn_kern_4_1}
\end{center}
\end{figure}

In Figure \ref{fig:knn_kern_4_1}, we observe that knn estimators for different $k$ oscillates for the samples smaller than $10000$ in the scenario \ref{ex_41}. However, it started to converge when $2000$ samples were used. MMD approximation also seems converged as well with the sufficient amount of samples. In Table \ref{tab:knn_rt_4_1}, it is obvious that the run time increases gradually for MMD approximation as the number of samples in use increases in the both scenarios \ref{ex_41} and \ref{ex_42}. When using knn estimator, there is insignificant increase in runtime as we use more samples for scenario \ref{ex_41} unlike MMD. Hence, knn estimator works poorly for the scenario 4.2 since it did not converge to a specific value as we increase the number of samples for the estimation.

\begin{table}[h!]
\centering
\resizebox{\columnwidth}{!}{%
\begin{tabular}{|l|l|l|l|l|l|l|l|l|l|l|}
\hline
 & 100 & 500 & $10^3$ & $2 \times 10^{3}$ & $3 \times 10^{3}$ & $4 \times 10^{3}$ & $5 \times 10^{3}$ & $7 \times 10^{3}$ & $10^{4}$ & $2 \times 10^{4}$ \\ \hline
\begin{tabular}[c]{@{}l@{}}mmd with Gaussian \\ for scenario 4.1\end{tabular} & $0.0094$ & $0.1024$ & $0.4214$ & $2.0664$ & $4.0022$ & $7.3511$ & $11.6014$ & $22.8216$ & $47.6443$ & $193.6788$ \\ \hline
\begin{tabular}[c]{@{}l@{}}knn (for all k) \\ for scenario 4.1\end{tabular} & $0.0030$ & $0.0084$ & $0.0186$ & $0.0301$  & $0.0450$ & $0.0598$ & $0.0700$ & $0.1002$ & $0.1511$ & $0.2941$  \\ \hline%
\begin{tabular}[c]{@{}l@{}}mmd with Gaussian \\ for scenario 4.2\end{tabular} & $0.0062$ & $0.1013$ & $0.4119$ & $2.2072$ & $4.7782$ & $8.4296$ & $15.4711$ & $25.5461$ & $50.4929$ & $198.4727$ \\ \hline
\end{tabular}
}
\caption{\scriptsize Average run times for scenarios \ref{ex_41} and \ref{ex_42} in seconds. Rows represent the approximations in \ref{knn_apprx} and \ref{kern_mmd} and columns represent the number of samples used.}
\label{tab:knn_rt_4_1}
\end{table}

\begin{figure}[h!]
\begin{center}
\includegraphics[width=8.4cm]{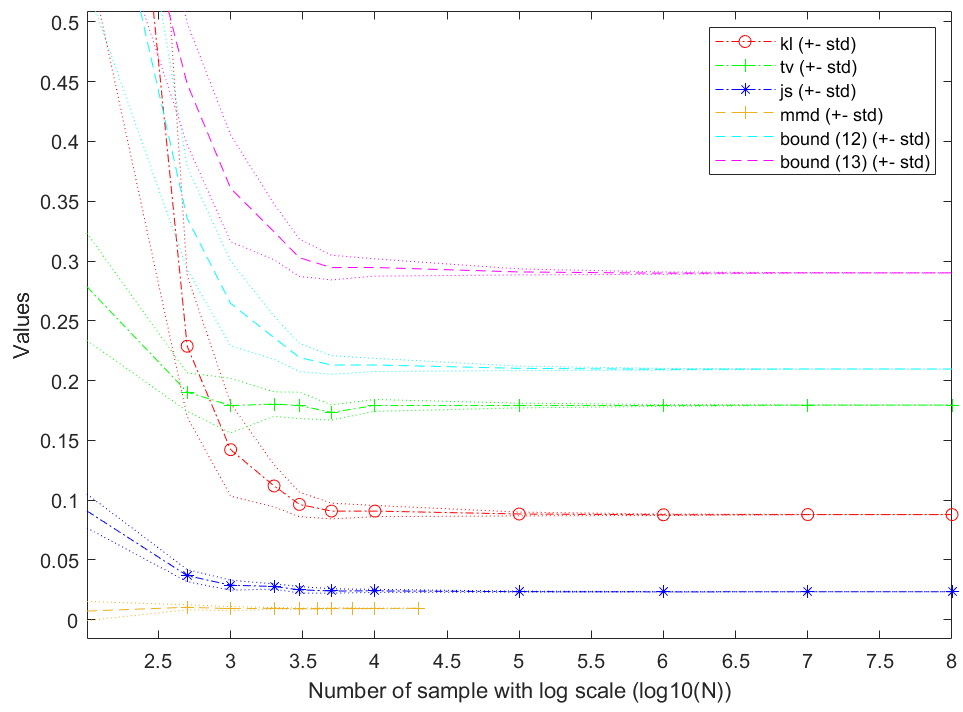} 
\caption{\scriptsize Average values of metrics for scenario 4.2 with respect to logarithm of the number of samples. Red line is KL, the green line is total variation  and the blue line is Jensen-Shannon divergences from \ref{leak_histo}. Dotted lines represent the standard deviation for each metric. Cyan line is from bound (\ref{eq:TV_KL_bound1}) and magenta line is from bound (\ref{eq:TV_KL_bound2}). Dark yellow line is MMD approximation from \ref{kern_mmd}} 
\label{fig:div_4_2}
\end{center}
\end{figure}

In Figure \ref{fig:div_4_2}, we observe for the scenario described in Section \ref{ex_42} that the convergence of approximations started to emerge when we use $10^5$ number of samples for histogram based estimations of the KL, TV and JS divergences. Standard deviation is quite tight after such a number of samples used. Since we do not know the explicit value of KL, we plug in the estimate into the bounds in \eqref{eq:TV_KL_bound1} and \eqref{eq:TV_KL_bound2}, which are shown in the figure as well. MMD approximation could only be possible to run with $20000$ samples maximum but it is sufficient for its convergence. Table \ref{tab:div_rt_4_2} shows that there is not a significant change in run times regardless of the number of samples used. 

\begin{table}[h!]
\centering
\resizebox{\columnwidth}{!}{%
\begin{tabular}{|l|l|l|l|l|l|l|l|l|l|l|l|}
\hline
 & 100 & 500 & $10^3$ & $2 \times 10^{3}$ & $3 \times 10^{3}$ & $5 \times 10^{3}$ & $10^{4}$ & $10^{5}$ & $10^{6}$ & $10^{7}$ & $10^{8}$ \\ \hline
\begin{tabular}[c]{@{}l@{}}kl  with \\ histogram\end{tabular} & $0.218$ & $0.21$ & $0.22$ & $0.21$ & $0.22$ & $0.23$ & $0.22$ & $0.21$ & $0.21$ & $0.26$ & $0.30$ \\ \hline
\begin{tabular}[c]{@{}l@{}}tv with\\ histogram\end{tabular}   & $0.034$ & $0.03$ & $0.03$ & $0.03$ & $0.03$ & $0.03$ & $0.03$ & $0.03$ & $0.03$ & $0.03$ & $0.03$ \\ \hline
\begin{tabular}[c]{@{}l@{}}js with \\ histogram\end{tabular}  & $0.331$ & $0.34$ & $0.33$ & $0.34$ & $0.33$ & $0.35$ & $0.33$ & $0.32$ & $0.33$ & $0.35$ & $0.42$ \\ \hline
\end{tabular}%
}
\caption{\scriptsize Average run times for the scenario in 4.2 in seconds. Rows represent the approximations in \ref{leak_histo} and columns represent the number of samples used.}
\label{tab:div_rt_4_2}
\end{table}



We were not able to run the LP calculation and the Sinkhorn's approximation for $W_1$ in the scenario \ref{ex_42} when using both histogram-based evaluations due to a memory problem. This indicates that the use of LP and Sinkhorn is impractical when the dimension of the samples are higher than 5. This seems to be due to the high amount of bins needed for the histograms in higher dimensions, which make the $C$ matrix too large with respect to memory.  

\section{Conclusion and Future Work}

To sum up, we evaluate the possible divergences and metrics to measure the mutual information $I(X;Y)$ for specific scenarios using multiparty computation in this paper. Results show that the histogram-based estimators of the divergences are strong for approximating the mutual information in terms of the number of samples used and the run time of the approximation. The MMD metric is also a useful measure for its convergence but it can be computationally heavy for applications requiring a high number of samples. Wasserstein distance is a quite informative metric as well. On the other hand, calculating it using LP or estimating it with Sinkhorn's algorithm becomes useless for the samples in the $5$ dimension.

\bibliography{ifacconf} 

\end{document}